\documentclass[12pt,comsoc]{IEEEtran}

\usepackage{amsmath}
\usepackage{amssymb}
\usepackage{graphicx}
\usepackage{cite}
\usepackage{bm}
\usepackage{dsfont}
\usepackage{graphics} 
\usepackage{caption}
\usepackage{subcaption}
\usepackage{epsfig} 
\usepackage{stackrel}
\usepackage{float}
\usepackage{soul}
\usepackage{tabu}
\usepackage{booktabs}
\usepackage{multirow}
\usepackage{array}
\usepackage[colorlinks,citecolor=blue]{hyperref}

\usepackage{xcolor}

\usepackage{ifthen}
\usepackage{xcolor,colortbl}

\definecolor{Gray}{gray}{0.85}
\definecolor{LightCyan}{rgb}{0.88,1,1}
\definecolor{grannysmithapple}{rgb}{0.66, 0.89, 0.63}
\definecolor{green(pigment)}{rgb}{0.0, 0.65, 0.31}

\title{Securing your Airspace: Detection of Drones Trespassing Protected Areas} 
\author{Alireza Famili, Angelos Stavrou, Haining Wang, Jung-Min (Jerry) Park, Ryan Gerdes}
\date{}

\begin{document}
	
\maketitle

\begin{abstract}
There has been a rapid growth in the deployment of Unmanned Aerial Vehicles (UAVs) in various applications ranging from vital safety-of-life such as surveillance and reconnaissance at nuclear power plants to entertainment and hobby applications. While popular, drones can pose serious security threats that can be unintentional or intentional. Thus, there is an urgent need for real-time accurate detection and classification of drones. In this article, we perform a survey of drone detection approaches presenting their advantages and limitations. We analyze detection techniques that employ radars, acoustic and optical sensors, and emitted radio frequency (RF) signals. We compare their performance, accuracy, and cost, concluding that combining multiple sensing modalities might be the path forward.
\end{abstract}

\section{INTRODUCTION}
Unmanned aerial vehicles (UAVs) have evolved rapidly over the past few decades leading to mass production of affordable drones~\cite{Matthan}. From kids and hobbyists to police officers and firefighters, drones have found novel applications and use cases. For instance, Google and Amazon use drones for merchandise delivery while law enforcement leverages drones for speed checks. During disasters, drones can help first responders to establish communications and locate victims. Unfortunately, similarly to most technological advancements, drones can be (ab)used for illicit activities~\cite{Matthan}. Indeed, criminals groups use drones to smuggle goods and breach secure locations, to name a few. Even benign uses of drones can be unlawful, including unintentional invasion of privacy, harm to humans and infrastructure due to collisions, and interference with other flying objects (e.g., airplanes). For example, in $2016$ Dubai airport reported that it had to shut down three times to avoid unauthorized drone activity~\cite{Matthan}. 

\begin{figure}
	\centering
	\includegraphics[height=2.8in,width=3.5in,trim={2.39cm 9.5cm 6.5cm 9.5cm},clip]{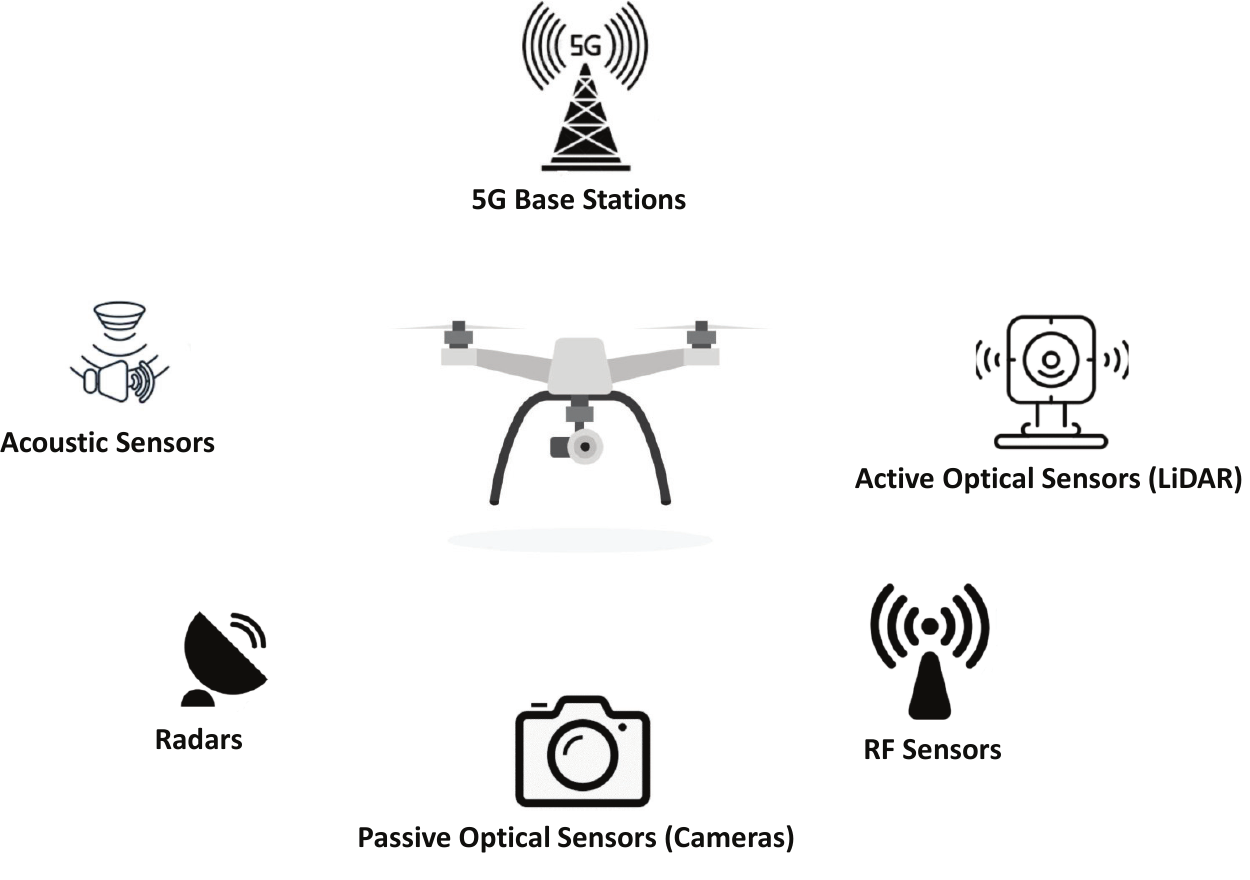} 
	\caption{Different drone detection technologies.}
	\label{Detection}
\end{figure}

Given the proliferation of drones and their use for both good and bad, there is an urgent need for accurate drone detection and classification into permitted and not-allowed while tracking their trajectory. Currently, there are different methods for detecting drones in the airspace: radars (e.g., \cite{mmWave_1}), acoustic sensors (e.g., \cite{9_substitute}), Radio Frequency (RF) signal detection (e.g., \cite{RF_substitute_1}), visual and optical sensors (e.g., \cite{LiDAR}), as shown in Fig.~\ref{Detection}. The drone detection system is typically deployed in close proximity to the area of interest. When the drone's trajectory enters or is projected to enter the no-fly zone, the detection system can track the drone and identify it either as a friendly or an unknown intruder. Subsequently, the system can notify an operator or enforce an automated policy.

In this survey article, we present an overview of the available approaches for detecting drones. Our aim is to understand the design space for drone detection techniques and expose any inherent or situational limitations for each of these approaches. We also explore other aspects that are pertinent to selecting the drone detection approach, including cost, power consumption, accuracy, and environmental variables that might affect the performance of the detection system. Initially, we present different radar approaches. We put a significant focus of the survey on radars because they are of the most promising methods in terms of accuracy. However, their cost and deployment requirements can render radars unsuitable for some use cases. Continuing to explore drone detection techniques, we discuss off-the-shelf acoustic sensors that offer a cost-effective but less accurate alternative for radars in some deployment scenarios. Next, we explore approaches based on RF transmission of the drone followed by visual and optical sensor detection methods. Finally, we end our survey with a discussion of multi-modal and sensor-fusion approaches in which multiple sensors are employed to improve the detection accuracy either in tandem or in sequence. A first-glance summary of the advantages and disadvantages of the drone detection approaches we cover in this article is summarized in Table~\ref{table:Comparison}.

\section{Radars}
The current state-of-the-art in moving object detection problem, whether it is detecting a big drone or a small bird, involves some form of radar. Evaluating the feasibility of using radar for UAV detection has received considerable attention~\cite{mmWave_1,5G_1,5G_2,3G}. However, there are some practical limitations and cost considerations when designing and deploying a radar suitable for detecting drones. 

\subsection{Radar Cross Section}
The main challenge is the variable size of UAVs which can make them invisible to traditional radars. Due to the small size of some UAVs and their main body construction in terms of materials, which can have a low reflection index, the Radar Cross Section (RCS) for them is extremely small and makes them hard to detect. Both theory and practice support this fact because the received power from the target object is a function of its RCS, and with smaller RCS, there will be smaller received power. 

\subsection{Frequency and Bandwidth}
Another design parameter for radars is their operating frequency: high-frequency radars are more expensive but can detect smaller UAVs. An additional advantage of high-frequency radars is their larger bandwidth, and finer resolution generates more accurate results. For example, a radar with a bandwidth of $1$GHz has a range resolution of $15$cm (range resolution equals the speed of light divided by twice the bandwidth).

\subsection{Radar Signal Power}
In practice, radars with higher transmission power offer improved detection results. In terms of wave modulation, $CW$ (Continues Wave) radars require significantly less power than pulsed versions and, thus, $CW$ radars are more attractive for small UAV detection. Many of the available research papers (e.g., \cite{mmWave_1}) use $FMCW$ (Frequency Modulated Continues Wave) radars for drone detection due to their lower power consumption compared to other wave modulation approaches.

\subsection{Active or Passive}
There are two different types of radar: active and passive. Active radars are equipped with both a transmitter and a receiver. The transmitter emits electromagnetic waves, which illuminates proximal targets. The receiver captures all reflected signals, which are then post-processed to expose any potential new targets. When only passive sensing is employed, the radar system is reduced to only receivers. Target illumination in the passive radar scenario is done by other signal sources, and the passive radar analyses the backscatter signals, including cellular signals, FM radio signals, WiFi signals, among others~\cite{3G}.

Although active sensing achieves a higher range of detection and better reliability, it requires significantly more transmit power and might not be capable of illuminating targets under diverse environmental conditions. Moreover, the radar operator needs to obtain a license and maintain permits for the band that the radar transmitter signal occupies. Passive radars do not require any operational permits because they do not actively transmit signals. Furthermore, their power consumption and cost requirements are significantly lower, accommodating multiple receivers for the same budget for a single active radar deployment. For instance, Chadwick et al.~\cite{3G} proposed a system for drone detection using passive radar technology leveraging available $UMTS \ 3G$ cellular communication signals as illumination sources. They considered three different ways for the illumination: using a cell phone on a call in the target area, having micro base stations for $3G$ communication in the target area, using the base stations in the closest vicinity of the target area. They use two receivers, one for capturing the genuine signal before all the reflections, and the other receiver is responsible for getting reflected signals. While this passive radar solution is cost-effective, it comes at the expense of accuracy and lack of reliable coverage.

\subsection{Beam Steering}
The more focused and narrow the transmitted signal, the better the illumination for detecting small objects. For instance, using omni-directional antennas with a wide main lobe will result in poor performance for detecting small objects. On the other hand, using a narrow radar beam with a focused main lobe, while accurate for small objects, it decreases the surveillance perimeter. One option is to use several antennas on the transmitter side, each with a narrow beam but placed in such a formation that, combined, they cover the target area. Another option is to make the transmitter mobile by using a rotor. This method is called mechanical beam steering, and it can cover the target area over a period of time. In addition to the mechanical beam steering, there is yet another approach called electrical beam steering in which a narrow transmitted beam scans small areas by changing the phase of the signal over time, resulting in full target area coverage.

\subsection{Mechanical or Multi-channel Scanning}
Like the transmitter beam steering for scanning the target area, the receiver can also be installed on a motor that can mechanically turn and scan the whole area. An alternative design is to be equipped with static multi-channel antennas that can receive signals from any direction. Most active radar scenarios where the transmitter and receiver are bundled together are usually either static multi-channel or use a mechanical rotation for both reception and transmission of signals. As an example, in \cite{mmWave_1}, Noetel et al. investigate two methods of scanning. In the first scenario, they used a scanning surveillance radar system which is a mechanically scanning $FMCW$ system operating in $94$~GHz (they used mmWave radar). This radar can scan with the frequency of $8$~Hz resulting in an image update rate of $8$ frames per second. In the second scenario, they used a multi-channel radar for perimeter surveillance which is static. However, the radar is equipped with four channels on the receiver side to cover the whole area. It is also able to determine the $3D$ location of the target. The multi-channel approach can be used in situations where mechanical scanning is prohibited. In both of the scenarios, since they used $FMCW$ radar, the power consumption is low. In addition, they were able to achieve good visibility of small objects and range resolution of $15$~cm because of the $1$~GHz bandwidth supported by the mmWave radar.

\subsection{Micro-Doppler Analysis}
\begin{figure}
     \centering
     \begin{subfigure}[b]{0.49\textwidth}
         \centering
         \includegraphics[width=\textwidth]{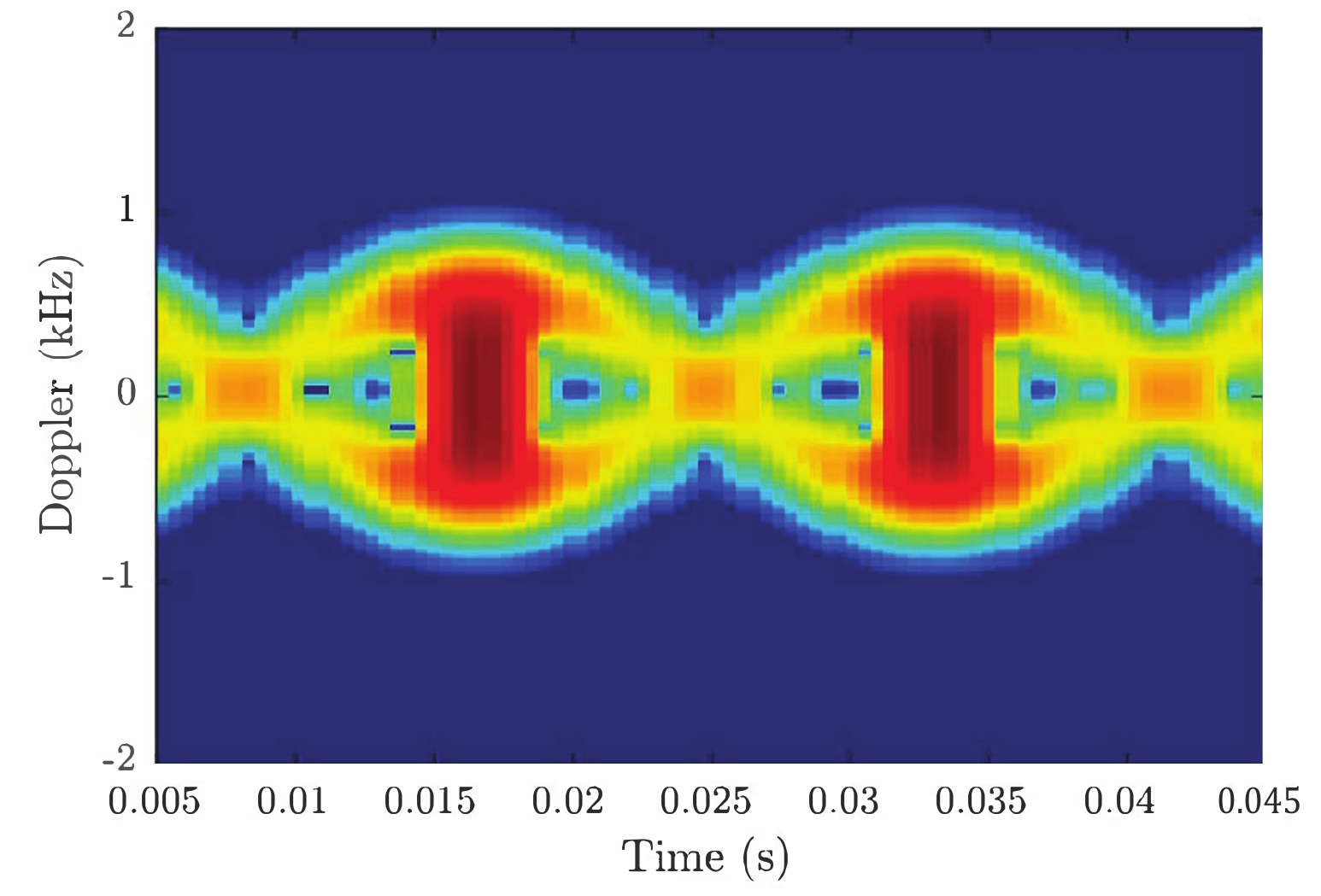}
         \caption{Blade length = $17$~cm}
         \label{microDoppler_17}
     \end{subfigure}
     \hfill
     \begin{subfigure}[b]{0.48\textwidth}
         \centering
         \includegraphics[width=\textwidth]{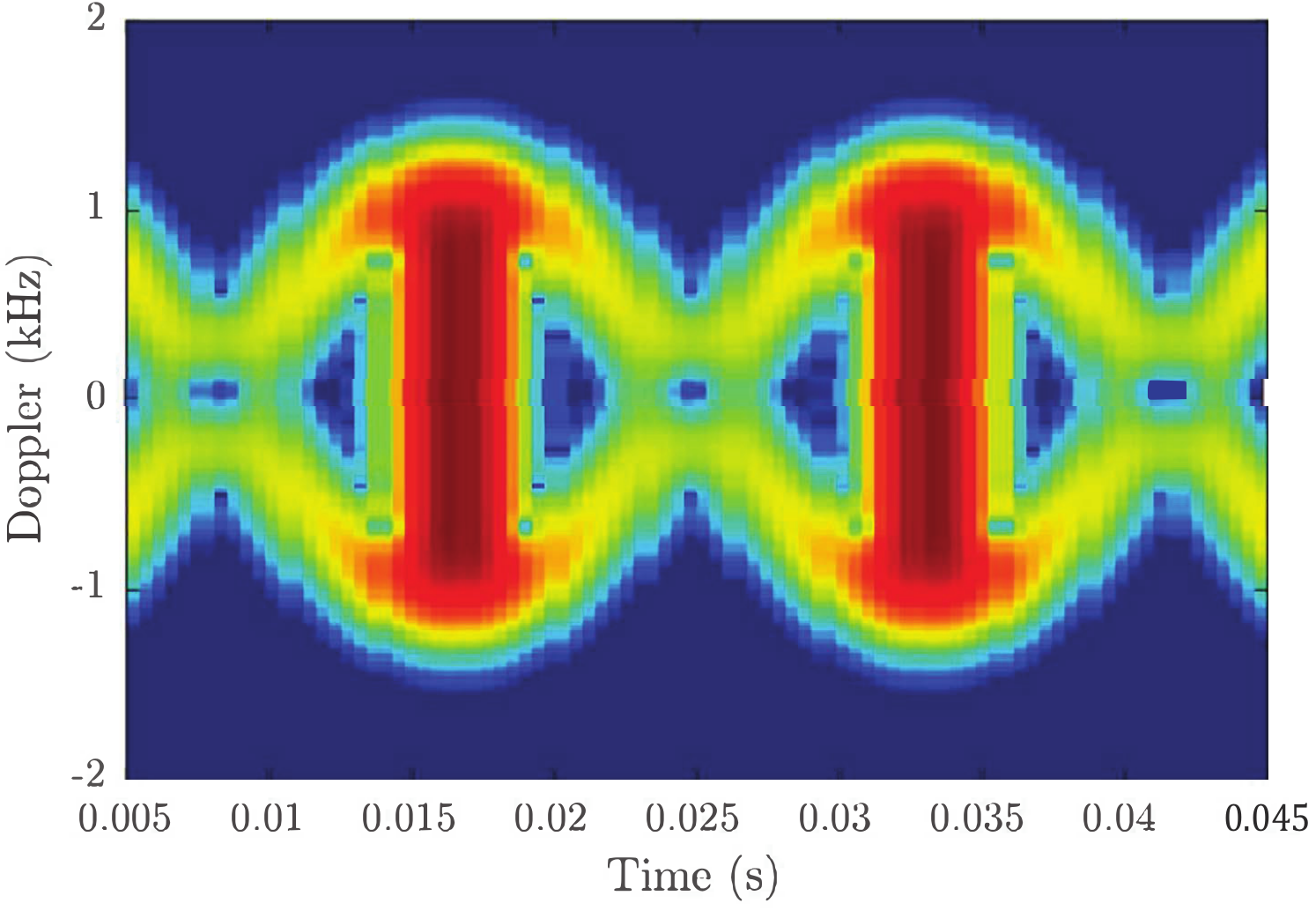}
         \caption{Blade length = $34$~cm}
         \label{microDoppler_34}
     \end{subfigure}
        \caption{Micro-Doppler analysis simulation: (a) Two $17$~cm blades (b) Two $34$~cm blades, rotating at $30$~Hz RPM which captured by a CW radar with center frequency of $2.41$~GHz~\cite{7_substitute}.}
        \label{microDoppler_17_34}
\end{figure}

 Micro-Doppler analysis is used in radar analysis to fingerprint and identify target objects. This is different than the Doppler effect used to determine the speed and direction of the target object. Any vibration or movement in the target object's body or any other moving parts on-board the target can be measured using micro-Doppler analysis~\cite{6_substitute}. Drone propellers are a good example of micro-Doppler analysis performed on the reflected radar signals from drones. In this case, micro-Doppler analysis can assist in distinguishing between drones and birds, reducing false alarms. In addition, using micro-Doppler analysis, we can estimate parameters related to the target drone, such as the length of the rotors' blades~\cite{7_substitute}. For example, in Fig.~\ref{microDoppler_17_34}, Gannon et al.~\cite{7_substitute} illustrated that when the size of a propeller's blades increases from $17$~cm to $34$~cm while maintaining the same rotation frequency of $30$~Hz, the Doppler response is doubled. 

\subsection{Future Radar Drone Detection}
One promising research direction is to leverage $5G$ cellular communications for drone detection. Terrestrial and satellite $5G$ communications can be used as both passive or active radar sources to illuminate and detect drones. Numerous research studies aim to investigate the challenges and limitations of harnessing existing and future $5G$ infrastructure capabilities for drone detection. As an example, Solomitckii et al.~\cite{5G_1} explored the idea of using $5G$ base station antennas for drone detection. Since $5G$ can employ mmWave antennas in the base station for communications, it is conceivable that the $5G$ infrastructure can also be used as radar for detection purposes. In addition, Wang et al.~\cite{5G_2} presented successful experimental results for drone detection from antennas operating in $28$~GHz, which are similar to the frequencies $5G$ base station antennas use.

We discussed the challenges and potential design parameters when selecting an appropriate radar for detecting small UAVs. We want to reiterate that the size and materials of the target UAVs, operating environment limitations, type of radar systems, and their associated cost of operation and deployment are the primary solution drivers when it comes to drone detection. 
Recent advances in radar technologies, including high range coverage, working in all weather environments, and continuous coverage capability during day and night, have elevated radar technologies as one of the best candidates for drone detection systems. In the following sections, we demonstrate how radar sensors can be used alone or in combination with other sensors to increase the reliability of flying object detection.

\section{Acoustic Sensors}
When it comes to drone detection, acoustic antennas can sense the unique noise produced by the drone's propellers and detect the presence of a drone in the area of interest. Some examples of the available research and literature that focus on drone detection using acoustic sensors can be found in~\cite{9_substitute} and \cite{8_substitute}. 

While economical, acoustic sensors have some significant drawbacks that need to be considered when it comes to drone detection. The primary limitation of acoustic sensors is that their performance is highly dependent on the range (distance) of the target to the sensor. Unfortunately, acoustic sensors are only suitable for short ranges, and they cannot be used to monitor large areas. To make matters worse, in crowded and noisy urban environments polluted with ambient sounds and noise, the performance of acoustic sensors degrades drastically. This limits their deployment options as acoustic sensors perform poorly in detection scenarios where patrol drones or other noisy equipment are employed to conduct the surveillance. In such cases, the ambient noise is too high, rendering any acoustic detection of external signals extremely hard to accomplish. On the plus side, acoustic sensors are inexpensive and can be easily acquired, installed, and deployed. In addition, they can perform well in any weather situation, both in the daytime or at night, and they do not need Line Of Sight (LOS) to the target object. All being said, when used on their own, they do not offer performance guarantees (due to their mentioned drawbacks); however, as a companion sensor, they can boost the overall system performance and accuracy.

\section{RF Ground Communication Sensors}
One of the most widely used approaches to detect the presence of a drone in no-fly zones is by sensing the RF communication between the drone and ground controllers. This method leverages RF sensors working as receivers scanning for RF communication channel transmissions. The RF sensors are designed to detect the RF frequency ranges that drones use for control and data signaling with ground controllers. The first step is to distinguish existing versus new RF communications. Then, using the newly extracted RF communications, they have to further identify unique RF signatures for drones using techniques such as the ones presented in~\cite{Matthan,RF_substitute_1,RF-substitute_2,Reviewer1_2}.

As an example, Nemer et al.~\cite{Reviewer1_2} propose a detection system based on the RF signature of drones which leverages ensemble learning to improve accuracy. Initially, a pre-processing stage removes the interference caused by other devices operating at the same frequency band as drones. Then, a machine learning program detects the presence of drone and identifies its type by deploying four classifiers hierarchically. 

For RF sensing of drones, the common assumption across all approaches is that there exists an RF communication link between the target drone and its ground controller. It is further assumed that this control signal can be captured and precisely analyzed even in the presence of other signals. Indeed, for many commercial drones, RF signals are the primary means for communicating navigation commands to the drone from the ground controllers and, reversely, when downloading captured data such as images, videos, and other sensory information captured by the drones. While these assumptions are valid for many commercial off-the-shelf drones, there are drones capable of flying autonomously without the need to receive periodic navigation commands. Moreover, in some scenarios, drones are equipped with an adequate amount of on-board memory to capture sensory information for prolonged periods of time. Thus, even when a drone supports RF communications, there could be extended periods of time in which there is no RF communication between drones and ground controllers. Another challenge with RF sensing for drone detection is the presence of environmental RF noise. This is especially true in urban areas where wireless activity is prevalent, generating overlapping and constant RF transmissions emanating from both ground and aerial targets that are not necessarily drones. For instance, people use their WiFi devices to stream videos from the Internet while they are walking on the high floors of a tall building resembling the movement and transmission originating from a drone. Thus, merely depending on RF sensing is not reliable for urban environments due to environmental and noise considerations, including the presence of multiple concurrent communications from both stationary and moving targets. On the other hand, in less populated or rural areas where there are few wireless devices, the RF channels are primarily silent. Therefore, it is easy to sense the communications between drones and their ground controller.

While using RF ground communication signals for drone detection has limitations, it offers a cost-efficient and easy-to-implement mechanism that can be useful when allowed to operate over a more extended period of time and in combination with other sensing modalities. Moreover, this type of RF sensing can work in any weather conditions, it is not affected by the time of day, and it does not need direct LOS to the target. Additionally, this method can detect the drone even before it takes off and when it appears to be stationary (i.e., the drone has landed or it is just hovering). As long as there exists an active RF communication link between the target drone and any ground controller, the RF sensors can detect it. More importantly, this is the only method that can locate the ground controller of the drone as well as the drone itself.

\section{Optical Sensors}
Using optical sensors, including cameras, gated lasers, and other visual sensing modalities that perform optical processing offers another means to detect and classify UAVs. Similar to using radars, there are two approaches for deploying optical sensors: active or passive. When using active sensing, the detection system leverages an optical signal that can be emitted by a gated laser (e.g., LiDAR) to illuminate an area or a target of interest. The detection occurs by processing the reflected optical signals from the target. The passive method leverages an optical receptor, such as a camera to capture images or video for visual processing and classification of drones. The main advantage of using cameras for drone detection is that visual processing of the image from a target can reveal additional useful information for its classification. Image and video processing techniques can be applied to distinguish between drones and other flying objects or birds, between intruder and friendly drones, and determine whether a drone is carrying explosives or weapons. Thus, visual sensing can go beyond mere object detection to object classification with high accuracy when available.

The major drawback of optical sensors is their dependence on an uninhibited LOS to the target. Moreover, their accuracy degrades significantly in visually impaired environments. For instance, even when using night vision cameras, the quality of captured information in reduced or deprived light settings is far from optimal. In fact, cameras may fail to produce reliable detection results for small targets under different weather conditions (e.g., foggy, cloudy, rainy, etc). Another issue is that cameras offer a narrow beam for detection which means that single cameras cannot cover the large areas of interest at once. Therefore, we have to use multiple cameras or rotate one camera to swipe the area of interest. While active visual sensing (i.e., lasers) are not as sensitive as regular cameras to weather conditions, they can only provide detection at a very short range from the target. Hammer et al.~\cite{LiDAR}, conducted experimental tests to evaluate the feasibility and practical performance of employing LiDAR for drone detection systems. While the results appear to be encouraging, the system had to operate in a very short range requiring a direct LOS to the target. When the target was within the sensor LOS and at a short range to the LiDAR system, a full $3D$ scan of the target was produced.

\section{Multi-Sensor Approach}
All of the sensor modalities discussed thus far have both advantages and limitations that can render them ideal or unreliable under certain environmental and weather conditions. We posit that a robust drone detection system should rely on more than one sensing modality. Carefully selected, multiple sensing modalities can complement each other, increasing the reliability and identification robustness of the overall system. Therefore, any drone detection system's perimeter surveillance and reliability can be enhanced by fusing the inputs from multiple sensors while optimizing their utility based on environmental conditions. 

For instance, Laurenzis et al.~\cite{multi_sensor_1} collected data from a heterogeneous sensor network consisting of acoustic antennas, small FMCW radar systems, and optical sensors. The authors applied acoustic sensors, radar, and LiDAR to monitor a wide azimuthal area ($360$ degree) to simultaneously track multiple drones with various degrees of success. In addition, they deployed optical sensors for sequential identification with a very narrow field of view. In another example~\cite{multi_sensor_static_moving}, Giovanneschi et al. propose a drone detection system that consists of two stations: one was a static multi-sensory network, and the other one was a sensor unit installed on-board a moving vehicle. They initially studied a fixed multi-sensory network that included an acoustic antenna array, a stationary FMCW radar, and a passive/active optical sensor unit. The active optical sensor was LiDAR. A mobile vehicle equipped with passive/active optical sensing was brought in to augment the sensory network and cover areas behind obstacles. In contrast, the static multi-sensory network monitored a stationary area with a sensor-dependent sensing coverage. The data fusion from the multi-sensory network and the moving vehicle provided an increased situational awareness for target detection.

	\begin{table*} [t]
	\scriptsize
		\caption{Comparison of drone detection approaches.}
		\label{table:Comparison}
		\begin{center}
			\begin{tabular} {|m{0.1\textwidth}|m{0.4\textwidth}|m{0.4\textwidth}|}
				\hline \rowcolor{green(pigment)} \textbf{Detection Techniques} & \begin{center} \textbf{Advantages} \end{center} & \begin{center} \textbf{Disadvantages} \end{center}\\
				\hline \rowcolor{grannysmithapple}
				\textbf{Radar} & Operates in Day/Night, Acoustic Noisy, and Visual Impaired environments. Long range. Constant tracking. Performs even if drone flies autonomously (without RF emissions). Configurable Frequencies, FMCW/Pulse, Static/Dynamic, Active/Passive. Drone size/type detection via micro-Doppler analysis.
				& Small RCS can affect performance (in cmWave radars, for a regular size commercial drones, RCS typically ranges from $-1$~dBsm to $-18$~dBsm). Active radars require transmission license and frequency check to prevent interference.\\
				\hline 
				\textbf{Acoustic microphone array} & Operates independent of visual conditions (day, night, fog, etc). Performs even if drone flies autonomously (without RF emissions). Operates in LOS/NLOS. Low cost implementation. Low energy consumption. Highly mobile and quickly deployable. & Short detection range (detection range $<$ 500~m). Performance degrades in loud and noisy environments. May work poorly in crowded urban environments due to acoustic noise. \\
				\hline \rowcolor{grannysmithapple}
				\textbf{RF signals}  & Operates in Day/Night, Acoustic Noisy, Visual Impaired, and LOS/NLOS environments. No licence required. Low cost sensors (e.g., SDRs). Can locate the controller of the drone on the ground. & Detection fails in cases where there is no RF signal transmission from the drone. May work poorly in crowded urban areas due to RF interference. \\
				\hline 
				\textbf{Vision-based}  & Offers ancillary information to classify the exact type of drone. Can record images as forensic evidence for use in eventual prosecution. & Short detection range (e.g., LiDAR sensors' range $<$ 50~m). Requires LOS. Relative expensive sensors. High computational cost. Performance degrades under different weather conditions (fog, dust, clouds, etc) and in visually impaired environments. \\
				\hline \rowcolor{grannysmithapple}
				\textbf{Multi-sensor} & Combines advantages of multiple different methods. Has better performance, higher accuracy, and long range detection. Robust under different scenarios and environmental conditions. & Increased cost and computational complexity compared to single sensors. \\
				\hline
			\end{tabular}
		\end{center}
	\end{table*}
	
\section{Conclusion \& Future Work Discussion} \label{conclusion}
We presented a survey on the available methods for drone detection. First, we covered the radar sensors with a deeper focus, as they are of the most promising approaches for detecting drones, but they come at a relatively high cost. In addition, we explored the capabilities and limitations of acoustic sensors due to their low energy requirements and deployment cost, showing that their use cases are limited to low noise environments. Moreover, we presented RF sensing as means for drone detection. This approach depends on detecting the drone's communications with a ground controller. Of course, drones might fly autonomously and remain silent for a prolonged period of time, thus preventing RF sensing from detecting their presence. Then we studied optical sensors which can be used actively, such as LiDAR, or in passive mode, like video and still imaging. Visual sensors offer advantages when it comes to target identification. However, they can be impaired by distance, lack of LOS to the target, and environmental conditions. Finally, we presented recent studies that attempt to fuse different sensing modalities to develop a more reliable approach for drone detection.

As we presented in this survey, using multiple types of sensors can mitigate some of the individual sensor limitations and boost detection robustness under adverse operational scenarios. Therefore, for future study, some may need to investigate how to alleviate the limitations of individual sensors by improving their performance using additional techniques (such as different learning algorithms for RF signature training) or by fusing them in a multi-sensor approach.

\footnotesize

\bibliographystyle{IEEEtran}
\bibliography{ref_Drone_Detection_Survey_arXiv_v0}


\section*{Biographies}
\textbf{Alireza Famili} (afamili@vt.edu) received his Bachelor’s and Master's degree in Electrical Engineering from the University of Tehran, Iran in 2016 and Lehigh University, USA in 2018, respectively. Alireza is currently pursuing a Ph.D. degree in electrical engineering at the Bradley Department of Electrical and Computer Engineering, Virginia Tech, USA.  His research interests include security issues in unmanned aerial vehicles (UAVs) and wireless communications/networking.

\textbf{Dr. Angelos Stavrou} is a Professor at the Bradley Department of Electrical \& Computer Engineering. 
Stavrou received his M.Sc. in Electrical Engineering, M.Phil. and Ph.D. (with distinction) in Computer Science all from Columbia University. 
His current research interests include security and reliability for distributed systems, security principles for virtualization, and anonymity with a focus on building and deploying large-scale systemss.

\textbf{Dr. Haining Wang} (hnw@vt.edu)  received his PhD from University of Michigan in 2003,  and he is currently a Professor in the Bradley Electrical and Computer Engineering Department at Virginia Tech. His research is focused on network and cloud security, especially intrusion detection and secure IoT systems.
He is an IEEE Fellow.

\textbf{Dr. Jung-Min (Jerry) Park} (jungmin@vt.edu) is a Professor and a Bradley Endowed Senior Faculty Fellow in the Department of Electrical and Computer Engineering at Virginia Tech.  Park is currently on leave from Virginia Tech, and is working at Samsung Electronics as a senior VP.  
Park’s research interests include dynamic spectrum sharing, emerging wireless technologies, and wireless security and privacy.  
Park is a recipient of several awards, including a 2017 Virginia Tech College of Engineering Dean’s Award for Research Excellence, a 2015 Cisco Faculty Research Award, a 2008 NSF Career Award, and a 2008 Hoeber Excellence in Research Award.  
He is an IEEE Fellow.

\textbf{Dr.\ Ryan M.\ Gerdes} (rgerdes@vt.edu) is an Associate Professor in the Bradley Department of Electrical and Computer Engineering at Virginia Tech.  He received his PhD.\ in electrical engineering from Iowa State University in August 2011. 
His research interests include cyber-physical systems security, with an emphasis on the operation of autonomous systems in unknown, uncertain, and adversarial environments; device fingerprinting, embedded systems security; sensor security; controls security; and cybersecurity.

\end{document}